# Research on the Quantum confinement of Carriers in the Type-I Quantum Wells Structure

Xinxin Li [1,2,5], Zhen Deng [1,2,4,5], Yang Jiang [1,2], Chunhua Du [1,2,4], Haiqiang Jia [1,2,3], Wenxin Wang [1,2,3] and Hong Chen [1,2,3,4,*]

[1]Key Laboratory for Renewable Energy, Beijing Key Laboratory for New Energy Materials and Devices, Beijing National Laboratory for Condensed Matter Physics, Institute of Physics, Chinese Academy of Sciences, Beijing 100190, People's Republic of China.
[2]Center of Materials and Optoelectronics Engineering, University of Chinese Academy of Sciences, Beijing 100049, People's Republic of China.
[3]Songshan Lake Materials Laboratory, Dongguan 523808, People's Republic of China.
[4]The Yangtze River Delta Physics Research Center, Liyang 213000, People's Republic of China.

[5]These authors contributed equally: Xinxin Li, Zhen Deng.
*Corresponding author. Email: hchen@iphy.ac.cn

## Abstract

Quantum confinement is recognized to be an inherent property in low-dimensional structures. Traditionally it is believed that the carriers trapped within the well cannot escape due to the discrete energy levels. However, our previous research has revealed efficient carrier escape in low-dimensional structures, contradicting this conventional understanding. In this study, we review the energy band structure of quantum wells considering it as a superposition of the bulk material dispersion and quantization energy dispersion resulting from the quantum confinement across the whole Brillouin zone. By accounting for all wave vectors, we obtain a certain distribution of carrier energy at each quantization energy level, giving rise to the energy subbands. These results enable carriers to escape from the well under the influence of an electric field. Additionally, we have compiled a comprehensive summary of various energy band scenarios in quantum well structures, relevant to carrier transport. Such a new interpretation holds significant value in deepening our comprehension of low-dimensional energy bands, discovering new physical phenomena, and designing novel devices with superior performance.

**Introduction**

Quantum confinement is considered to be an inherent property in low-dimensional material structures, which has attracted enormous attention in the fields of optoelectronics[1, 2], microelectronics[3, 4], and quantum physics[5, 6]. Typically, the quantum confinement effect refers to the phenomenon where the behavior of particles is influenced by quantum mechanics. Such kinds of effects appear when the carriers are confined in a length scale comparable to or smaller than the de Broglie wavelength, resulting in discrete levels of energy [7]. The unique electronic energy band structure because of the quantum confinement effect leads to many novel physical and optical properties [8-10], which are utilized for devices with better performance. For example, quantum tunneling diodes with tunneling current as the main current component have been applied in low-power and high-speed circuits due to their efficient electronic transmission [11-13]. In addition, the study of quantum confinement effects has also promoted the establishment of band theory for low-dimensional materials, which reveals the essence of carriers' motion in the microscopic world with the quantum theory concept. Therefore, band analysis has become a powerful tool not only for understanding material properties but also for predicting the physical and chemical properties of the system to achieve better device performance [14-17].

After years of in-depth research, we have experimentally observed the great carrier extraction efficiency (more than 85%) in InGaN Multiple quantum wells (MQWs) [18-20], InGaAs MQWs[21, 22], and other low-dimensional materials systems [23] in PIN structures, as shown in Table 1. Such a high carrier extraction efficiency in low-dimensional systems is exciting, especially in the case of thick barrier (>10 nm) and high potential. Moreover, the carrier extraction efficiency can be further improved (over 90%) by applying reverse bias. These phenomena indicate that the efficient carrier extraction phenomenon in PIN structure does not originate from thermal excitation [24] or tunneling [25] as suggested in studies. The NIN structures with the same low-dimensional structure have also been studied in the same way, but high extraction efficiency did not occur as that in PIN structures because of the carrier blocking in the well [26]. However, in classical views, it is believed that the photo-generated carriers in the low-dimensional structures cannot escape from the potential well to form the photocurrent but can only relax to the bottom of the ground state and then recombine due to the discrete energy levels [27], known as the quantum confinement of carriers. Obviously, the traditional energy band theory is no longer suitable for analyzing the anomalous photo-generated carriers' extraction in the QWs structure under the PN junction. Since the properties of the MQWs system are often understood through the band structures, it is necessary to go back to reanalyze the energy band structure in order to identify the essence of the phenomenon.

Table.1 The great carrier extraction efficiency in the low-dimensional PIN structures of our previous researches.

| Structures | Active region | Well thickness (nm) | Well energy gap ($E_g$) | Barrier Thickness (nm) | Barrier energy gap ($E_g$) | Extraction efficiency (%) (@0 V) | Reference |
|---|---|---|---|---|---|---|---|
| Quantum wells | (InGaAs/GaAs) × 10 | 5 | ~1.2 | 20 | 1.43 | 87.3 | [21] |
| | (InGaN/GaN) × 10 | 2.5 | 2.7 | 14 | 3.4 | 96.5 | [19] |
| | (InGaN/GaN) × 10 | 2.5 | 2.7 | 12 | 3.4 | 95.15 | [18] |
| Quantum dots | InAs/GaAs | diameter: 20 height: 6 | 0.41 | 50 | 1.43 | 88 | [23] |

In this paper, the energy band structure of the QWs is reanalyzed and considered as the superposition of the dispersion of bulk material and the dispersion of quantization energy from the quantum confinement within the whole Brillouin zone, rather than the discrete energy levels at a certain vector given in traditional band theory. There is a certain energy distribution of the carrier's energy band in real space, making it possible for carriers to escape from the QW under the electric field. Moreover, various cases for energy bands of the Type-I QW structures have been summarized for carrier transportation, confirming the high possibility of carrier escape, which provides a good explanation for the efficient carrier extraction found in previous studies. This understanding of the energy bands can serve as a supplementary explanation to the traditional energy band theory of low-dimensional quantum confinement systems.

The MQWs structure is composed of alternating semiconductor thin layers with different bandgaps, as shown in figure 1(a), with a growth direction along the z-axis. When the thickness of one layer (blue) is comparable to the electron's de Broglie wavelength, the wave function that could originally propagate freely is confined by the other materials on both sides, resulting in the quantum confinement of carriers in z direction. The main feature of MQWs structure is that the barrier thickness is thicker so that the interaction between the QWs can be ignored, and the structure can be considered as a periodic repetition of a single QW [27]. Therefore, when analyzing the energy band structure of MQWs structure, we can simplify it as analyzing a single QW's energy band. Here, the energy dispersions of the well and barrier are shown in figure 1(b) for a schematic explanation of energy alignment. The energy band gaps of the well and the barrier are $E_{g1}$ and $E_{g2}$, while the blue and yellow lines correspond to energy dispersions of the well and the barrier layers, respectively. For the equilibrium heterojunction (the QW structure), the chemical potential determines the energy band arrangement of the well and barrier based on the phase equilibrium conditions [28], resulting in the discontinuities in the conduction and valence bands as shown in figure 1(b). It is this discontinuity or the so-called band offset that plays a crucial role in confining electrons or holes in a QW structure leading to the quantum behavior [29]. Different from the discrete energy levels at a certain vector given in the traditional band theory, the real energy dispersion of a QW structure should be a combination of the dispersion of the material itself and additional energy elevation caused by the quantum confinement effect within the whole Brillouin zone.

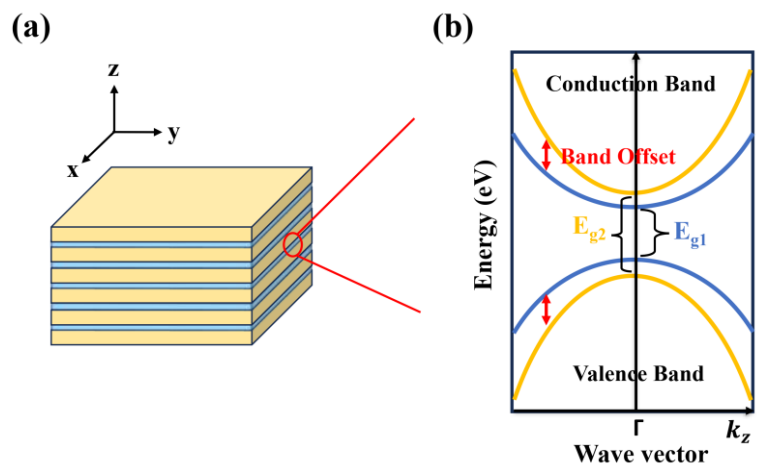

Figure 1. Schematic diagrams of the structure, and band structure of QWs. (a) The typical illustration of the MQW structure. (b) The schematic diagram of the energy dispersions of the well and barrier in a single QW in $k$-space.

As analyzed above, the energy $E_z$ along the z direction of the carriers in the well consists of two parts:

$$E_z = E_z^b + E_z^s \tag{1}$$

where the $E_z^b$ represents the energy resulting from the interaction of crystal cells while the $E_z^s$ is the quantization energy caused by quantum confinement.

The methods for calculation of energy $E_z^b$ in bulk materials include the Tight Binding Approximation method (TBA) [30], the Orthogonal Plane-Wave method (OPW), the k.p method [31], and the pseudopotential method [32], etc. The main process is to solve the single electron Schrödinger equation in a periodic external field, which is not illustrated in detail here. As for the quantization energy $E_z^s$ caused by the quantum confinement, the band offset is one of the key parameters determining energy level lifting. Assuming that the QW is an ideal square potential well that the quantization energy $E_z^s$ can be obtained by solving Schrödinger equation in the one-dimensional potential well with finite depth. The potential value is the band offset obtained from the difference of $E_z^b$ dispersions (see figure 1(b)). Thus, Schrödinger equation for stationary states in the position basis in the well is given as [33]:

$$-\frac{\hbar^2}{2m^*_{(\vec{k}_z)}}\frac{d^2}{dz^2}\psi_{E_z^s(\vec{k}_z)}(z) + V_{(\vec{k}_z)}(z)\psi_{E_z^s(\vec{k}_z)}(z) = E_z^s{}_{(\vec{k}_z)}\psi_{E_z^s(\vec{k}_z)}(z), \tag{2}$$

where $m^*_{(\vec{k}_z)}$ is the effective mass of the carrier, $z$ is the position, $\vec{k}_z$ is the wave vector, $\psi_{E_z^s(\vec{k}_z)}(z)$ is the wave function of the particle corresponding to the quantization energy

$E_{z(\vec{k}_z)}^s$, and $V_{(\vec{k}_z)}(z)$ is the well potential generated by the band offset.

The Schrödinger equation presented here aims to solve the quantization energy levels at all vectors. In this scenario, the effective mass $m^*_{(\vec{k}_z)}$ varies with different $\vec{k}_z$ vectors [34], as does the well-potential $V_{(\vec{k}_z)}$ due to the varying band offsets [35].

$$m^*_{(\vec{k}_z)} = \begin{cases} m^*_{w(\vec{k}_z)} = \dfrac{\hbar^2}{\left(\dfrac{d^2 E_z^{b-w}{}_{(\vec{k}_z)}}{d\vec{k}_z^2}\right)} & |z| < \dfrac{a}{2} \\ m^*_{b(\vec{k}_z)} = \dfrac{\hbar^2}{\left(\dfrac{d^2 E_z^{b-b}{}_{(\vec{k}_z)}}{d\vec{k}_z^2}\right)} & \dfrac{l}{2}|z| \geq \dfrac{a}{2} \end{cases} \quad (3)$$

$$V_{(\vec{k}_z)}(z) = \begin{cases} 0 & |z| < \dfrac{a}{2} \\ V_{(\vec{k}_z)} = E_z^{b-b}{}_{(\vec{k}_z)} - E_z^{b-w}{}_{(\vec{k}_z)} & \dfrac{l}{2}|z| \geq \dfrac{a}{2} \end{cases} \quad (4)$$

Where $m^*_{w(\vec{k}_z)}$ is the effective mass of the carrier in the well, $m^*_{b(\vec{k}_z)}$ is the effective mass of the carrier in the barrier, $E_z^{b-w}$ is the energy dispersion in the well, $E_z^{b-b}{}_{(\vec{k}_z)}$ is the energy dispersion in the barrier, $a$ is the well thickness, and $l$ is the period thickness of the QW.

The method for solving the above Equation 2 of different $\vec{k}_z$ vectors consists of writing the general forms of the wave functions inside and outside the well, and then matching the values and slopes at the boundary to solve for the unknown constants. Assuming that $0 \leq E_{z(\vec{k}_z)}^s < V_{(\vec{k}_z)}$, the general solutions of Equation 2 are [35]:

$$\psi_{E_z^s(\vec{k}_z)} = A \sin \alpha z + B \cos \alpha z \quad |z| < \dfrac{a}{2} \quad (5)$$

$$\psi_{E_z^s(\vec{k}_z)} = C e^{-\beta z} + D e^{\beta z} \quad \dfrac{l}{2} > |z| \geq \dfrac{a}{2} \quad (6)$$

where $\alpha = \dfrac{\sqrt{2 E_{z(\vec{k}_z)}^s m^*_{w(\vec{k}_z)}}}{\hbar}$, $\beta = \dfrac{\sqrt{2(V_{(\vec{k}_z)} - E_{z(\vec{k}_z)}^s) m^*_{b(\vec{k}_z)}}}{\hbar}$, A, B, C, and D are constants to

be determined.

Application of the boundary conditions (continuity and continuity of derivatives) at $|z| = \frac{a}{2}$ and $|z| = \frac{l}{2}$ gives the transcendental equations about β and α, and the energy spectrum of $E^s_{z(\vec{k}_z)}$ at different $\vec{k}_z$ vectors will be determined by the values of β and α.

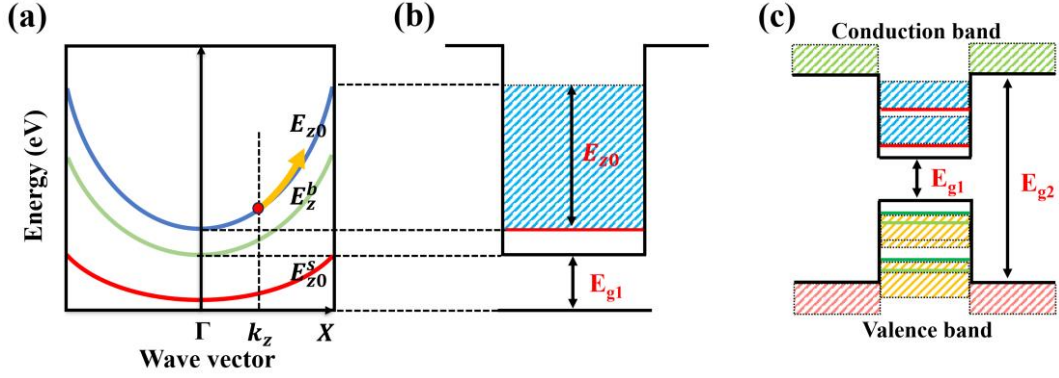

**Figure 2.** Schematic diagrams of real energy dispersion $E_z$. (a) The schematic diagram of energy dispersion $E_z$ consists of the energy dispersion ($E^b_z$) of the bulk material and dispersion of the quantum energy level $E^s_z$ due to quantum confinement with all the vectors. (b) The energy subband structure in the real space for electrons from the energy dispersion $E_{z0}$ in (a). (c) The whole energy band structure in the real space for carriers in the QW structure.

To better understand the real energy $E_z$ in the QW structure as mentioned above, in particular to clarify the mechanisms of how the carriers effectively run out of the well, a schematic diagram is drawn in figure 2(a). The green line represents the energy dispersion of the bulk material ($E^b_z$) along the z-direction, while the red line represents the dispersion of the quantization energy level due to quantum confinement (taking the ground state $E^s_{z0}$ as an example). When two types of energy are superimposed, the whole energy dispersion of $E_{z0}$ becomes the blue line in figure 2(a), which exhibits a certain range of energy in the whole Brillouin zone. In this case, there is a high probability for the carriers at the $k_z$ vector (the red point) transporting to higher energy along the dispersion curve (see the yellow arrow in figure 2(a)), as long as the additional required energy is provided. When an electric field exists, the carriers in the well can obtain the required energy and momentum for transportation through the continuous interaction with the lattice. While in classical band theory, the increase in carriers' energy under an electric field is unlikely to compensate for the large gap between the quantization energy levels at one time, leading to the failure of transportation. Furthermore, the dispersion $E_z$ in figure 2(a) is transferred into the real space for demonstration purposes, which only focuses on the distribution range of energy, as shown in figure 2(b). As we can see, the red lines in the well in figure 2(b) are the quantization energy level $E_{z0}(\Gamma)$ of electrons at $\Gamma$ point, while the blue shaded area represents the energy distribution of the $E_{z0}$ in real space appearing as a subband rising from the red energy level. The whole energy bands structure in the real space for carriers in the QW structure is given in figure 2(c), where the blue shaded area and the yellow shaded area represent the energy bands for electrons and holes in real space, respectively, and the green and red shaded areas are the energy bands in the barrier. One thing for sure is that for both electrons and holes in the well, there is a subband arises from each quantization energy (red lines for electrons, light green lines for light holes, and dark green lines for heavy holes) when considering all the vectors in the whole Brillouin zone. As long as the subband states in the well can extend above the barrier energy, there is a high possibility for carriers escaping from the well with the help of the electric field.

Table 2. The 16 permutations of energy bands for carriers in the QW structure. Where Y and N represent the conditions of the transportation described in figure 3(a) are met and not, respectively, while the red and black letters correspond to the electrons and holes, respectively. For example, the YNYN means that for both electrons and holes, the condition of transportations-1 is met while the condition of transportations-2 is not.

|      | NN   | NY   | YN   | YY   |
|------|------|------|------|------|
| **NN** | NNNN | NNNY | NNYN | NNYY |
| **NY** | NYNN | NYNY | NYYN | NYYY |
| **YN** | YNNN | YNNY | YNYN | YNYY |
| **YY** | YYNN | YYNY | YYYN | YYYY |

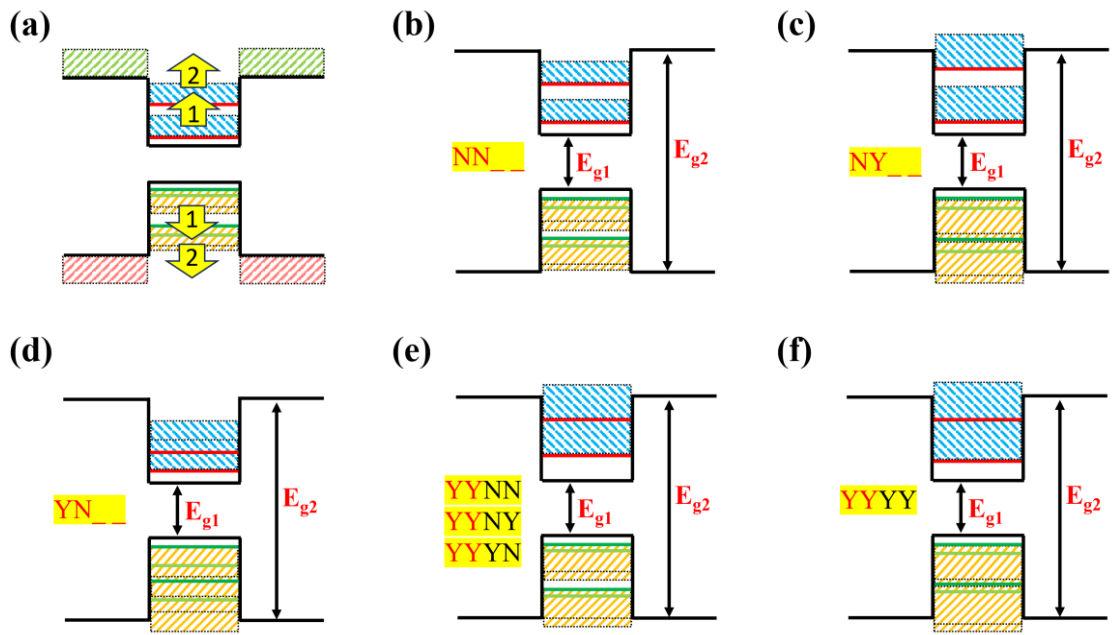

**Figure 3.** Multiple possibilities for energy bands in the QW structure assuming that there are two subbands in the well. (a) Two transportations of carriers (electrons or holes) in the well. Transportations-1: carriers transport from one subband to the other. Transportations-2: carriers transport from the higher subband in the well to the band in the barrier. (b)-(d) The energy bands for carriers in the QW structure corresponding to the permutations described in the first row (NN_ _), second row (NY_ _), and third row (YN_ _) of table 2, respectively. (e) The energy bands for carriers in the QW structure with permutations of YYNN、YYNY、YYYN described in table 2. (f) The energy bands for carriers in the QW structure with the permutation of YYYY described in table 2.

According to the energy bands for carriers in the QW structure shown in figure

2(b), we attempt to analyze the possible scenarios of the energy bands. The transportation of the carriers (electrons or holes) in the QW is shown in figure 3(a), assuming that there are two subbands in the well. Transportations-1 (Trans-1): there is no gap between the subbands in the well, allowing carriers to transport from one to another under an electric field. Transportations-2 (Trans-2): there is no gap between the higher subband in the well and the band in the barrier, allowing carriers to transport out of the well under the electric field. The energy bands for carriers in the QWs structure in figure 3(a) can be summarized into 16 permutations according to these transportations, as shown in table 2. The letters Y and N represent the conditions of the transportation described in figure 3(a) are met and not, respectively, while the red and black letters correspond to the electrons and holes, respectively. For example, the YNYN means that for both electrons and holes, the condition of Trans-1 is met while the condition of Trans-2 is not. For both electrons and holes, the conditions for carriers escaping from the well are very strict. Typically, only when both Trans-1 and Trans-2 are simultaneously fulfilled, can the carriers (electrons or holes) finally escape from the well. In addition, because of the existence of the Coulomb effect, both electrons and holes must fulfill the escape conditions at the same time to ensure the realization of carriers escape. Considering these limitations, the 16 mentioned permutations can be categorized into two cases: the carriers can escape (YYYY) or cannot (the others). Furthermore, the energy bands for carriers in the QW structure based on these 16 permutations are displayed with more intuitive diagrams in figures 3(b)-3(f). As shown in figures 3(b)-3(d) (NN_ _, NY_ _, and YN_ _), there are energy gaps in the well due to the small widths of the electrons' subbands that the Trans-1 and Trans-2 for electrons cannot be allowed simultaneously, resulting in the confinement of electrons. In this situation, carriers cannot escape from the well due to Coulomb interaction whatever the energy bands of the holes are. Similarly, when the electrons can escape but the holes are still confined in the well, effective carrier escape still cannot occur, as exhibited in figure 3(e). Only when both electrons and holes can be extracted simultaneously, can the effective escape of carriers be achieved, as displayed in figure 3(f). Unfortunately, even if the conditions of carrier escape are harsh, the probability of the YYYY permutation is very high since the conduction band and valance are usually very wide. Our discussion provides a direction to find out the quantum confined condition for carriers in the QW structure.

The band structure is an important semiconductor parameter, shaping numerous physical and chemical properties of low dimensional structures represented by quantum wells, such as electrical, optical, physicochemical, and other characteristics. Consequently, any alterations in the band structure inevitably result in modifications to the internal properties of the structures. Such changes could prompt a renewed comprehension of quantum effects and necessitate reinterpretation of experimental results. Furthermore, with this new comprehension of band structure in low-dimensional structure, it is expected to achieve special devices with novel characteristics under the strong built-in electric field of the p-n junction and promote technological innovation in semiconductor optoelectronic conversion application fields.

In conclusion, after reanalyzing the energy bands of the MQWs structure, we believe that in the confinement direction, the real energy $E_z$ of the carriers consists of two parts, one is the dispersion $E_z^b$ of the bulk material and the other one is the dispersion $E_z^s$ of quantization energy caused by quantum confinement within the whole Brillouin zone, rather than the discrete energy levels given at a certain vector in traditional band theory. The energy distribution of the carrier's energy leads to the band formation, making it possible for carriers to escape from the well under the electric field. Based on this, multiple possibilities for energy bands in the QW structure have been summarized for carrier transportation. Such supplementary explanation to the traditional band theory would significantly assist the discovery of new physics phenomena and the development of novel semiconductor devices.

## Reference

[1] Köhler R, Tredicucci A, Beltram F, Beere HE, Linfield E H, Davies A G, Ritchie D A, Iotti R C and Rossi F 2002 Terahertz semiconductor-heterostructure laser *Nature* **417** 156-9

[2] Wang H X, Fu Z L, Shao D X, Zhang Z Z, Wang C, Tan Z Y, Guo X G and Cao J C 2018 Broadband bias-tunable terahertz photodetector using asymmetric GaAs/AlGaAs step multi-quantum well *Appl. Phys. Lett.* **113** 171107

[3] Yang M J, Wang F C, Yang C H, Bennett B R and Do T Q 1996 A composite quantum well field-effect transistor. *Appl. Phys. Lett.* **69** 85-7

[4] Ferry D K, Weinbub J, Nedjalkov M and Selberherr S 2022 A review of quantum transport in field-effect transistors *Semicond. Sci. Technol.* **37** 043001

[5] Fu H L et al 2019 3/2 fractional quantum Hall plateau in confined two-dimensional electron gas *Nat. Commun.* **10** 4351

[6] Liu C X, Qi X L, Dai X, Fang Z and Zhang S C 2008 Quantum anomalous Hall effect in $Hg_{1-y}Mn_yTe$ quantum wells *Phys. Rev. Lett.* **101** 146802

[7]     Ashrafi A Quantum Confinement: An Ultimate Physics of Nanostructures In *Encyclopedia of Semiconductor Nanotechnology* Edition: 1 ed, Umar A Ed American Scientific Publishers: 2011, pp 1-67

[8]     Brum J A and Bastard G 1986 Resonant carrier capture by semiconductor quantum-well *Phys. Rev. B* **33** 1420-3

[9]     Barbagiovanni E G, Lockwood D J, Simpson P J and Goncharova L V 2014 Quantum confinement in Si and Ge nanostructures: Theory and experiment *Appl. Phys. Rev.* **1** 011302

[10]   Kalt H and Klingshirn C F 2019 *Semiconductor Optics 1: Linear Optical Properties of Semiconductors* 5 ed. (Chan, Springer International Publishing) pp 251-271

[11]   Day D J, Chung Y, Webb C, Eckstein J N, Xu J M and Sweeny M 1990 Double quantum-well resonant tunnel-diodes *Appl. Phys. Lett.* **57** 1260-1

[12]   Althib H 2021 Effect of quantum barrier width and quantum resonant tunneling through InGaN/GaN parabolic quantum well-LED structure on LED efficiency *Results Phys.* **22** 103943

[13]   Encomendero J, Yan R S, Verma A, Islam S M, Protasenko V, Rouvimov S, Fay P, Jena D and Xing H G 2018 Room temperature microwave oscillations in GaN/AlN resonant tunneling diodes with peak current densities up to 220 kA/cm$^2$ *Appl. Phys. Lett.* **112** 103101

[14]   Gu Y, Zhang Y G, Ma Y J, Zhou L, Chen X Y, Xi S P and Du B 2015 InP-based type-I quantum well lasers up to 2.9 μm at 230 K in pulsed mode on a metamorphic buffer *Appl. Phys. Lett.* **106** 121102

[15]   Yahyazadeh R 2021 Effect of hydrostatic pressure on the radiative current density of InGaN/GaN multiple quantum well light emitting diodes *Opt. Quantum Electron.* **53** 571

[16]   Xue J, Zhao Y J, Oh S H, Herrington W F, Speck J S, DenBaars S P, Nakamura S and Ram R J 2015 Thermally enhanced blue light-emitting diode *Appl. Phys. Lett.* **107** 121109

[17]   Chen B L 2017 Active region design and gain characteristics of InP-based dilute bismide type-II quantum wells for mid-IR lasers *IEEE Trans. Electron Devices* **64** 1606-11

[18]   Wu H Y et al 2016 Direct observation of the carrier transport process in InGaN quantum wells with a pn-junction *Chin. Phys. B* **25** 117803

[19]   Yang H J, Ma Z G, Jiang Y, Wu H Y, Zuo P, Zhao B, Jia H Q and Chen H 2017 The enhanced photo absorption and carrier transportation of InGaN/GaN Quantum Wells for photodiode detector

applications *Sci. Rep.* **7** 43357

[20] Li Y F et al 2018 Visualizing light-to-electricity conversion process in InGaN/GaN multi-quantum wells with a p-n junction *Chin. Phys. B* **27** 097104

[21] Sun Q L, Wang L, Jiang Y, Ma Z G, Wang W Q, Sun L, Wang W X, Jia H Q, Zhou J M and Chen H 2016 Direct observation of carrier transportation process in InGaAs/GaAs multiple quantum wells used for solar cells and photodetectors *Chin. Phys. Lett.* **33** 106801

[22] Liu J, Wang L, Sun L, Wang W Q, Wu H Y, Jiang Y, Ma Z G, Wang W X, Jia H Q and Chen H 2018 Anomalous light-to-electricity conversion of low dimensional semiconductor in p-n junction and interband transition quantum well infrared detector *Acta Phys. Sin.* **67** 128101

[23] Wang W Q et al 2016 Carrier transport in III-V quantum-dot structures for solar cells or photodetectors *Chin. Phys. B* **25** 097307

[24] Botha J Rand Leitch A W R 1994 Thermally activated carrier escapr mechanisms from $In_xGa_{1-x}As$/GaAs quantum-wells *Phys. Rev. B* **50** 18147-52

[25] Emiliani V, Bonanni B, Frova A, Capizzi M, Martelli F and Stone S S 1995 Tunneling and relaxation of photogenerated carriers in near-surface quantum wells *J. Appl. Phys.* **77** 5712-7

[26] Tang X S, Li X X, Yue C, Wang L, Deng Z, Jia H Q, Wang W X, Ji A C, Jiang Y and Chen H 2020 Research on photo-generated carriers escape in PIN and NIN structures with quantum wells *Appl. Phys. Express* **13** 071009

[27] Grundmann M 2006 *The Physics of Semiconductors* (Berlin, Springer)

[28] Ashcroft N and Mermin D 1976 *Solid State Physics* (Philadelphia, Saunders College) Vol 46

[29] Nag B R 2000 *Physics of Quantum Well Devices* 1ed. (Dordrecht, Kluwer Academic Publishers) pp 22-34

[30] Chadi D J and Cohen M L 1975 Tight-binding calculations of the valence bands of diamond and zincblende crystals *Phys. Status Solidi B-Basic Solid State Phys.* **68** 405-19

[31] Hermann C and Weisbuch C 1977 K.P perturbation theory in III-V compounds and alloys: a reexamination *Phys. Rev. B* **15** 823-33

[32] Schwerdtfeger P 2021 The Pseudopotential Approximation in Electronic Structure Theory *ChemPhysChem* **12** 3143-55

[33] Schrödinger E 1926 Quantisierung als Eigenwertproblem *Ann. Phys.* **384** 489-527

[34] Kittel C 1957 *Introduction to Solid State Physics* 2 ed. (New York, John Wiley & Sons Inc.) p 289

[35] Schiff L I 1968 *Quantum Mechanics* 3ed. (New York, McGraw-Hill Book Company) Vol 2 pp 19-44

**Acknowledgements**

This work was supported by the National Natural Science Foundation of China (Grant Nos. 61991441 and 62004218) the Strategic Priority Research Program of Chinese Academy of Sciences (Grant No. XDB01000000) and Youth Innovation Promotion Association Chinese Academy of Sciences (2021005). We also appreciate the financial supports from the Center for Clean Energy.

**Contributions**

Chen H conceived the study. Li X X performed the investigation and analysis with the assistance of Deng Z. Li X X and Deng Z wrote the text with continuous feedback from Chen H. All authors discussed the results and contributed to editing the manuscript.